\documentclass[aps,prd,preprintnumbers,showpacs]{revtex4}
\usepackage{graphicx}

\begin{document}

%
%

\eprint{Nisho-1-2023}
\title{A Way of Determination of Axion Mass with Quantum Hall Effect}
\author{Aiichi Iwazaki}
\affiliation{International Economics and Politics, Nishogakusha University,\\ 
6-16 3-bantyo Chiyoda Tokyo 102-8336, Japan }   
\date{Jun. 26, 2023}
\begin{abstract}
Axion dark matter is converted to electromagnetic radiations in the presence of strong magnetic field.
The radiations possibly give rise to non trivial phenomena in condensed matter physics.
Especially, we discuss that saturation of plateau-plateau transition width observed at low temperature in integer quantum Hall effect 
is caused by the axion. The radiations from axions are inevitably present in the experiment.
Although the radiations generated by axion is extremely weak, Hall conductivity jumps up to next plateau
even if only a single electron occupies an extended state; a localized electron 
is transited to the extended state by absorbing the radiation.  According to our analysis,
previous experiment\cite{sat6} of the saturation in detail suggests 
that the axion mass is in the range $10^{-5}\mbox{eV}\sim 10^{-6}$eV.
We propose a way of the determination of the axion mass by
imposing microwaves on Hall bar and also a way of the confirmation that
the axion really causes the saturation of the width.
\end{abstract}
\hspace*{0.3cm}

\hspace*{1cm}

\maketitle

\section{introduction}
One of most significant issues in particle physics is to find an evidence beyond the standard model.
Especially, finding axion is the important step exploring a new physics. It also gives rise to
a solution of dark matter in the Universe.
Axion is the Goldstone boson of Peccei Quinn symmetry\cite{axion1,axion2,axion3}, which naturally solves strong CP problem.
Such an axion is called as QCD axion. 
The allowed mass range\cite{Wil,Wil1,Wil2} of the QCD axion is severely restricted such as $m_a=10^{-6}\mbox{eV} \sim10^{-3}$ eV.
Hereafter we only consider QCD axion, although our argument may be applicable to non QCD axion, dark photon e.c.t.
In this paper we use the units, $c=1$, $k_B=1$ and $\hbar=1$.
 
Many projects\cite{admx,carrack,haystac,abracadabra,organ,madmax,brass,iwazaki01} for the detection of the dark matter axion
have been proposed 
and are undergoing at present. 
Most of them use a mechanism that the axion is converted to electromagnetic radiation 
under strong magnetic field. Namely, the axion dark matter produces electromagnetic fields when it is imposed
by strong magnetic field.  The radiations are detected with the use of
resonant cavity, Rydberg atom, LC circuit, e.t.c.  

\vspace{0.1cm}
The axion dark matter may cause some of astrophysical phenomena including gravitational lens, or
phenomena not yet well understood such as fast radio bursts\cite{fastradioburst1,fastradioburst2}.
The axion dark matter may also give rise to significant effects in condensed matter physics at low temperature.
Coherence of electrons in conductor, semiconductor, e.t.c. is in general disturbed by thermal effects.
When they are disconnected from thermal bath, their coherence is restored completely.
But, even with the disconnection, the electrons are in the bath of axion dark matter. Although the coupling of the axion
with individual electron is extremely small, the axion effect may be observed\cite{magnon,center} because a large number of electrons are involved
in condensed matter.

\vspace{0.1cm}
In this paper we consider axion effect in quantum Hall phenomena\cite{girvin}, which are realized in two dimensional electrons
under strong magnetic field $B$ perpendicular to them. 
We focus on integer quantum Hall effects. 
It is well known that
Hall resistance $\rho_{xy}$ ( conductance $\sigma_{xy}$ ) is quantized such as
$\rho_{xy}=(2\pi/e^2)\times 1/n$ ( $\sigma_{xy}=-\rho_{xy}^{-1}$ ) with positive integer $n$ and electric charge $-e$ of electron.
The diagram of $\rho_{xy}$ in variable $B$ 
shows plateaus\cite{von,aokiando,halperin} of $\rho_{xy}$ for each $n$. They look like steps, while $\rho_{xx}$ ( $\sigma_{xx}$ ) vanishes
on the plateaus. But $\rho_{xx}$ ( $\sigma_{xx}$ ) takes non zero values in small width $\Delta B$ between two plateaus.
That is, $\rho_{xy}$ increases larger than the value $(2\pi/e^2)\times 1/n$ on a plateau and forms next plateau $\rho_{xy}=(2\pi/e^2)\times 1/(n-1)$
as magnetic field $B$ increases. There is small width $\Delta B$ between the two plateaus. 
The width shrinks as temperature $T$ decreases.
In other wards, the plateau-plateau transition is not sharp. The plateau $ (2\pi/e^2)\times 1/n$
smoothly connects the plateaus $(2\pi/e^2)\times 1/(n-1)$ when temperature $T\neq 0$.
But, it is expected theoretically that at $T=0$ the plateau-plateau transition 
is sharp because $\Delta B(T)$ is expected to vanish at $T=0$.   
But actual experiments show that the width $\Delta B$ saturates a finite value as $T$ goes to zero.
Plateau-plateau transition is not sharp. It is generally believed that the saturation is intrinsic although
plausible explanation for the intrinsic is still lack. In this paper,
we discuss that axion dark matter causes the saturation. That is, axions are converted to electromagnetic radiations under
the strong magnetic field $B$. The radiations disturb coherence of electrons, which result non vanishing $\Delta B$ at $T=0$.
The radiations converted from axions play a role of external noise against the measurement of $\Delta B$ just like 
thermal noise.
( Sometimes, the derivative $d\rho_{xy}/dB$ is discussed to see the sharpness of the plateau-plateau transition. 
The saturation of $\Delta B\neq 0$ corresponds to
the saturation of $d\rho_{xy}/dB\neq \infty$ as $T \to 0$. )

\vspace{0.1cm}
Quantum Hall effect\cite{von} is realized in two dimensional electron system. The system is fabricated in MOSFET or heterostructures of
GaAs. The system shows various intriguing phenomena under strong magnetic field such as not only quantization of Hall resistance but also
Josephson-like effect\cite{joseph1,joseph2,joseph3} e.t.c..  Quantum Hall system has been extensively investigated since the discovery, 
but some of phenomena are not still fully understood. One of the phenomena is the saturation\cite{sat1,sat2,sat3,sat4,sat5,sat6} of the width
$\Delta B$ in plateau-plateau transition of integer quantum Hall effect.
The width arises at $T\neq 0$ because thermal effects cause localized electrons below Fermi energy $E_f$ to transit to
extended states above $E_f$. They occupying the extended states carry electric current so that
the value of electric Hall resistance decreases in the transition region.
We discuss that similar effect\cite{sat1,micro} to the thermal one may arise owing to electromagnetic radiations produced
by axion dark matter. The radiations are produced under the strong magnetic field, which is used in the experiment of quantum Hall effect.
The radiations make localized electrons transit to the extended states so that the non vanishing width $\Delta B$ arises even at $T=0$.
With imposing microwaves on Hall bar, we can determine the axion mass by observing the saturation of the width. 
Decreasing their frequencies $\nu$ of the microwaves, the width $\Delta B(\nu)$ decreases\cite{micro}. But
when the frequency is below a critical frequency, 
we observe the saturation. The critical frequency $\nu_c$ is given by the axion mass $m_a$; $\nu_c=m_a/2\pi$.

\vspace{0.1cm}
The detection of electron's absorption of radiations generated by axions has been performed with the use of Rydberg atoms.
In the experiment, Carrack\cite{carrack}, ionized atoms are detected so that a number of the atoms,
which absorb radiations, are needed for the detection.  In our case, there are many electrons $\sim 10^{8}$
in Hall bar with typical size $500\mu \mbox{m}\times 200\mu$m$=10^{-3}\rm cm^2$.  Furthermore, only a single electron is sufficient
to make Hall conductivity jump up to next plateau.

\vspace{0.1cm}
In next section(\ref{2}) we explain integer quantum Hall effect. In section(\ref{3}) 
in order to explain the presence of Hall plateau, we give a brief review of 
localization of electrons in integer quantum Hall effect.
In section(\ref{4}) we discuss how axion dark matter affects the width
$\Delta B$ in plateau-plateau transition. In section(\ref{5}) we give a brief review of the features of
QCD axion dark matter. We estimate transition probability of localized electron to extended states by absorbing 
radiations converted from axion in section(\ref{6}). In section(\ref{7}) we show the number of electrons steady present
in extended states, taking account of the inverse process that electrons occupying extended states transit back to
localized states by spontaneously emitting radiations. 
In section(\ref{8r}), we calculate the relation between $\Delta B$ and the width $\Delta E_f$ of Fermi energy,
within which the transition from plateau 
$(2\pi/e^2)\times 1/(n-1)$ to plateau $(2\pi/e^2)\times 1/n$ is complete. We find that the dependence agrees well to
experimental results. Finally, we propose a way of the determination of axion mass by imposing external microwaves to Hall bar.
Furthermore, we propose a way of the confirmation that axion really produces the saturation of the width $\Delta B$ in section(\ref{8}). 
Finally, conclusions follow in section({\ref{9}}).

\section{integer quantum hall effect}
\label{2}
We make a brief review of integer quantum Hall effect\cite{girvin}. The states of
two dimensional free electrons under magnetic field $B$ are specified by integer $n \geq 0$, so called Landau levels
with their energies specified as $E_n$ ).
( The two dimensional space is specified by the coordinates $x$ and $y$ and $\vec{B}=(0,0,B)$. 
The two dimensional electrons are formed by being trapped in a quantum well between for instance, GaAs and AlGaAs. 
The motion of the electrons in the direction $z$ is energetically forbidden, while the motion in the direction $x$ and $y$ is
energetically allowed. )
There are a number of degenerate states in each Landau level
with the degeneracy $eB/2\pi$; the number density of degenerate state in each Landau level.
( Density of states is described as a delta function $\propto \delta (E-E_n)$. )
Their wave functions are given by

\begin{equation}
\Phi(\vec{x},n,k)=C_{n}H_n\Big(\frac{x-l_B^2k}{l_B}\Big)\exp\Big(-\frac{(x-l_B^2k)^2}{2l_B^2}\Big)\exp(-iky)
\end{equation} 
with normalization constant $C_{n}=(2^{n}n!\sqrt{\pi}l_BL_y)^{-1/2}$,
where $L_y$ is the length of the Hall bar in the direction $y$ and $H_n(x)$ denotes Hermite polynomials. 
$l_B=\sqrt{1/eB}$ is called as magnetic length ( $l_B\simeq 1.8\times 10^{-7}\rm cm (B/10T)$ ).
We use a gauge potential $\vec{A}_{ex}=(0,Bx,0)$.
Their energies are $E_n=\omega_c( n+1/2)$ with cyclotron frequency $\omega_c=eB/m^{\ast}$; effective mass $m^{\ast}$ of electron
in semiconductor. Generally $m^{\ast}$ is much smaller than real mass $m_e\simeq 0.5$MeV of electron. ( For instance,
$m^{\ast}=0.067m_e$ in GaAs. Hereafter, we use physical parameters of GaAs. )   
Degenerate states in each Landau level are specified by $k$. 
Cyclotron frequency is of the order of $\sim 10^{-2}(B/10T)$eV.

\vspace{0.1cm}
Electron possesses spin components with up and down. Thus, 
each Landau level is split to two states with energies $E_{n\pm}=\omega_c(n+1/2)\pm g\mu_B B$ owing to Zeeman effect;
$\pm$ correspond to spin parallel and anti-parallel to the magnetic field $B$, respectively. Namely,
Zeeman energy $\pm g\mu_B B$ with $g\simeq 0.44$ and Bohr magneton $\mu_B=e/2m_e$.
Typically, Zeeman energy is of the order of $10^{-3}(B/10T)$eV.  
Obviously Zeeman energy is smaller than the cyclotron energy in semiconductor.
The degeneracy ( number density of electrons ) of each Landau level with definite spin is given by $eB/2\pi$.  
 
Starting from the lowest energy state, the states in higher Landau levels are occupied. 
Namely,
increasing the number density of electron $\rho_e$, higher Landau levels with larger energies are occupied. 
Usually, we use an index so called filling factor $\nu=2\pi\rho_e/eB$.
It implies for instance, Landau level with energy $E_{0-}$ is fully occupied but Landau level with $E_{0+}$ is
partially occupied when $2 > \nu >1$.

\vspace{0.1cm}
Flowing an electric current in $x$ direction of Hall bar, we measure voltages in $x$ and $y$ direction, i.e. $V_{xx}$ and $V_{xy}$, respectively.
Then we obtain $\rho_{xy}$ and $\rho_{xx}$.  
It is remarkable phenomena\cite{von,aokiando,halperin} that Hall resistance $\rho_{xy}=(2\pi/e^2)\times 1/n$ is constant even when 
the filling factor $\nu$ changes within a range $n+1>\nu>n$: Landau levels up to $n$ are fully occupied, while
the level with $n+1$ is partially occupied. That is, we see plateaus in the diagram of $\rho_{xy}$ in $B$. 
The plateaus arise owing to the localization of two dimensional electrons.

\section{localization of two dimensional electrons}
\label{3}
\begin{figure}[htp]
\centering
\includegraphics[width=0.6\hsize]{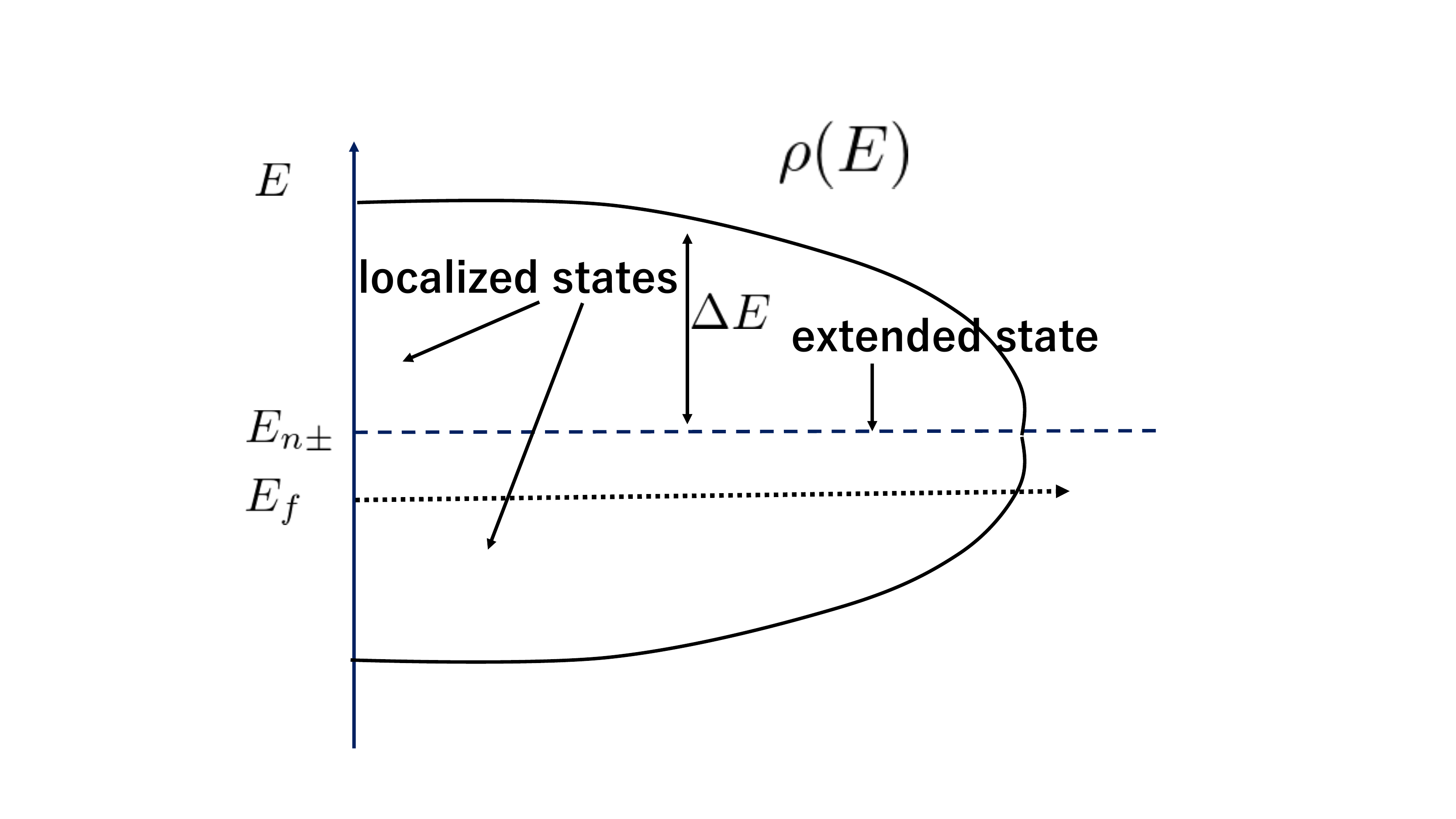}
\caption{Density of state $\rho(E)$. Dashed line denotes a density of state of free electrons under magnetic field $\vec{B}$.
Dotted line represents Fermi energy $E_f$}
\label{a}
\end{figure}

We need to explain localization of two dimension electrons in order to understand integer quantum Hall effect.
In actual materials, there are impurities, defects e.t.c. which lift up the degeneracy in Landau level. Electrons
do not freely move, but randomly move scattering with them. That is, electrons move in a random potential $V$ caused
by disorders such as impurities, defects, e.t.c. .
As a result, most of electrons are localized so that they
cannot carry electric currents. But, a small fraction of them are not localized so that they can carry electric current.
As is well known, all of two dimensional electrons are localized\cite{localization} around impurities, defects, e.c.t. when there is no external magnetic
field ( i.e. there is no interactions which break time reversal symmetry. ) 
Thus, even if magnetic field $B$ is present, it is naturally expected that most of electrons are localized,
although non-localized electrons are present. We stress that
localized electrons cannot carry electric currents, while non-localized electrons carry electric currents.
The point for
understanding integer quantum Hall effect is the presence of non-localized ( extended )\cite{aokiando,ono} electrons when $B\neq 0$.
Although the degeneracy is lifted up, there exist a small fraction of states with energy $E=E_{n\pm}$.
Electrons in such a state are not localized.   Actually, numerical simulations show that 
when there is a potential $V$ representing disorder, 
the density of states $\rho(E)$ has finite width around the energy $E_{n\pm}$
shown schematically in Fig.\ref{a}

The effect of the potential $V$ makes the density of state $\rho(E)$ to have finite width $\Delta E$ in the distribution.
It is assumed that the width is much less than the cyclotron frequency $\omega_c=eB/m^{\ast}$ in strong magnetic field $B$.
It holds for the potential energy which are much smaller than the cyclotron frequency. 
That is, $\Delta E \ll \omega_c$. ( $\omega_c\sim 10^{-2}$eV with $B\sim 10$T. ) 
For instance, $\rho(E)\propto \sqrt{1-((E-E_{n\pm})/\Delta E)^2}$ with $E-E_{n\pm}>\Delta E$\,\cite{andouemura}. 
Furthermore, the potential is supposed to have
the same amount of attractive and repulsive components. So, the form of $\rho(E)$ is symmetric around the center
$E_{n\pm}$ as shown in the figure.

\vspace{0.1cm}
When the variation length of the potential is much larger than the magnetic length $l_B\simeq 8\times 10^{-7}\rm (10\mbox{T}/B)^{1/2}$cm, 
we may consider that electrons move along the contour of the potential, circulating with small radius $l_B$. 
Because almost contours are closed, electrons moving along the contours
are localized. On the other hand, a small fraction of the contours extends from boundary to boundary. 
Such electrons moving along the extended contours
can carry electric current from boundary to boundary. It is the physical picture of localization.
Even if the variation length of the potential is of the same order of $l_B$, the numerical simulation\cite{simulation} shows that there exist extended states with 
their energies $E_{n\pm}$ and other states with energies $\neq E_{n\pm}$ are localized in the case of infinitely large Hall bar. 

\vspace{0.1cm}
When Fermi energy $E_f$ is less than $E_{n+}$ but larger than $E_{n-}$, electrons which can carry electric currents, are in
extended states whose energies are equal to $E_{n-}$ or $E_{m\pm}$ with $n>m$. 
For instance, for $E_{1+}>E_f>E_{1-}$,
Hall resistance is given such that
$\rho_{xy}=(2\pi/e^2)\times 1/3$, while Hall conductance $\sigma_{xy}$ is such that $\sigma_{xy}=-1/\rho_{xy}$
( Conventionally,  we refer Hall conductance as positive value like $\sigma_{xy}=n\times e^2/2\pi$. )
As long as the Fermi energy $E_f$ is in the range $E_{1+}>E_f>E_{1-}$, the Hall resistance does not vary
so that a plateau is formed.
On the plateau, we have longitudinal resistance $\rho_{xx}=0$,
which implies that the electric current flowing in $x$ direction is dissipationless. 
The currents are carried by electrons in extended states
with energies such as $E_{1-}$ and $E_{0\pm}$.  Namely,
when we add an electron, it occupies a localized state with energy $E_f <E_{1+}$
which does not carry electric current. Thus, Hall resistance does not change.
But, if the electron occupies the extended states with the energy $E_f=E_{1+}$, it carry electric current. Then, Hall resistance
suddenly down to $ \rho_{xy}=(2\pi/e^2)\times 1/4$, or Hall conductivity rises up to $\sigma_{xy}=4\times e^2/2\pi$
from $\sigma_{xy}=3\times e^2/2\pi$. The transition is sharp like the step function.

\vspace{0.1cm}
We should note that when we increase magnetic field $B$ without changing the number density of electrons,
the degeneracy $eB/2\pi$ increase but the Fermi energy decreases. The Fermi energy changes with magnetic field $B$.
Then, we see plateau in diagram of $\rho_{xy}$ vs $B$,
as long as $E_f$ does not cross the energy $E_{n\pm}$. Inversely when we decrease $B$, $E_f$ moves upward ( increases ).
If $E_f$ passes the energy, for instance, $E_{1+}$, Hall resistance
steps down from $\rho_{xy}= (2\pi/e^2)\times 1/3$ to  $\rho_{xy}= (2\pi/e^2)\times 1/4$. In other wards, $\sigma_{xy}$ steps up
from $\sigma_{xy}=3\times e^2/2\pi$ to  $\sigma_{xy}=4\times e^2/2\pi$. 
This is because the electrons occupying the states with the energy $E_{1+}$ ( $<E_f$ ) carry electric current.
It should be noticed that even if a single electron occupies the extended states with the energy $E_{1+}$, 
the Hall conductance jumps to $\sigma_{xy}=4\times e^2/2\pi$ from $\sigma_{xy}=3\times e^2/2\pi$.
This is a striking feature of quantum Hall effect. The feature is understood in topological argument\cite{topology1,topology2}.

Hereafter, we examine in detail the case of $E_{1-}< E_f <E_{1+}$ for concreteness.

\vspace{0.1cm}
Up to now, we have discussed localization in the infinite large size of Hall bar, in which 
extended states have the energies $E_{n\pm}$. But in the case of finite size,
there are localized states with their sizes larger than 
the size of Hall bar. Such states carry electric current and have energies in the range $E_{n\pm}-\delta \le E \le E_{n\pm}+\delta$
where $\delta(L_h)$ depends on the size $L_h$ of Hall bar such as $\delta (L_h\to \infty)=0$.
In general $\delta$ is smaller than the width $\Delta E$ of $\rho(E)$. 
We call such states as effective extended states because they can carry electric current.
Hence, when the Fermi energy $E_f$ in the range  $E_{n\pm}-\delta \le E_f \le E_{n\pm}+\delta$, 
for instance $E_{1+}-\delta \le E_f \le E_{1+}+\delta$,
Hall resistance $\rho_{xy}$ takes the value $(2\pi/e^2)\times 1/4$.

\section{axion effect on plateau-plateau transition}
\label{4}
\begin{figure}[htp]
\centering
\includegraphics[width=0.6\hsize]{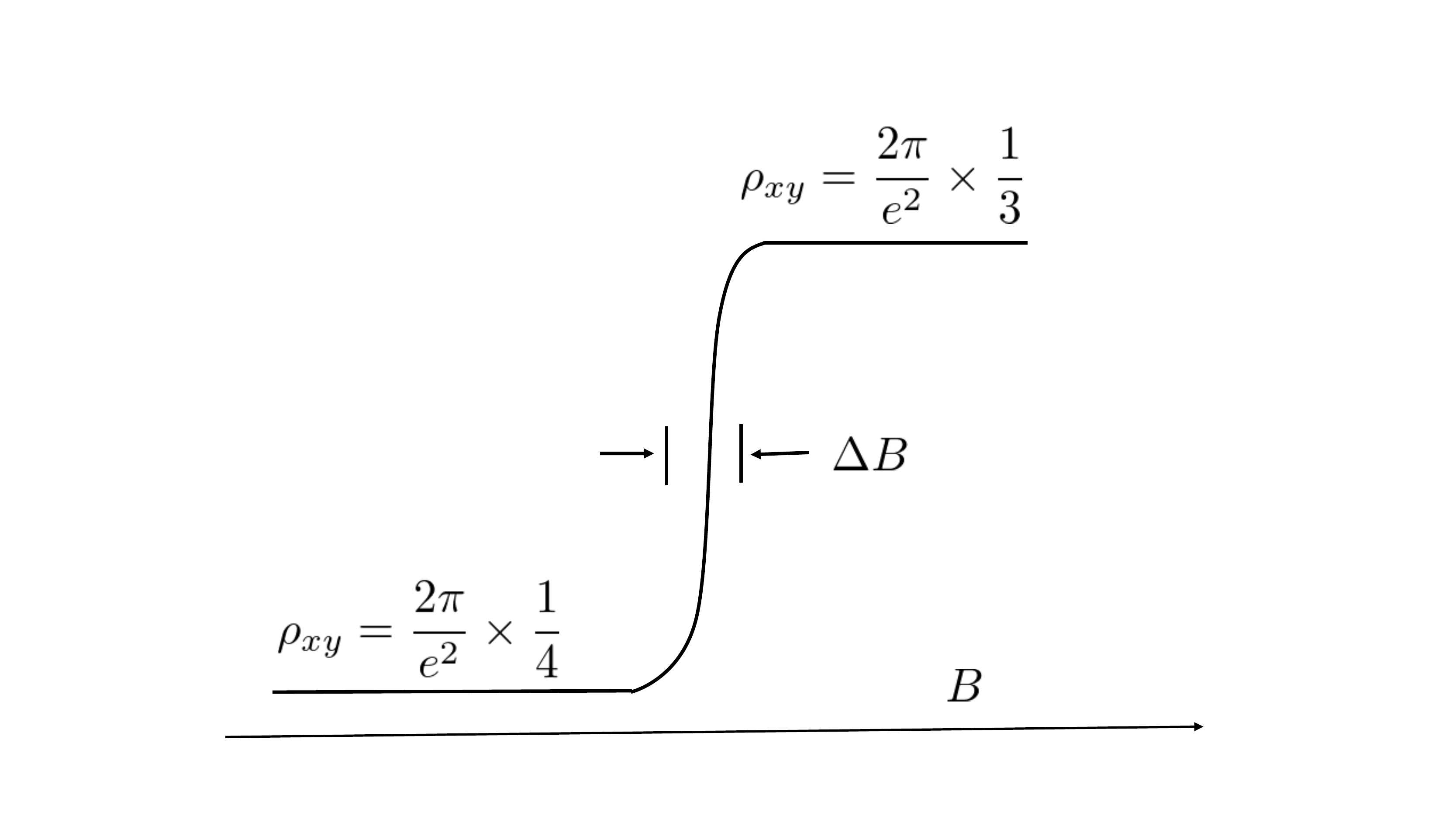}
\caption{plateau-plateau transition with width $\Delta B$}
\label{b}
\end{figure}

In the above explanation, we do not include thermal effect on the electrons. When temperature $T\neq 0$, probability that electrons
occupy the states with energies larger than Fermi energy does not vanish. 
It is proportional to $\exp(-|E-\mu(T)|/T)$ with chemical potential $\mu(T)$; $\mu(T)\to E_f$ as $T\to 0$.
It implies that even if $\mu(T)$ is less than $E_{1+}$,
there are non vanishing probability of electrons occupying the extended states with the energy $E_{1+}$.
So the Hall conductivity $\sigma_{xy}$ does not keep
the value $\sigma_{xy}=3\times e^2/2\pi$, becoming larger than this one. It
gradually increases toward the value $\sigma_{xy}=4\times e^2/2\pi$ when chemical potential $\mu$ approaches the energy $E_{1+}$.
( Inversely, Hall resistance $\rho_{xy}=2\pi/e^2\times 1/3$ goes down to $\rho_{xy}=2\pi/e^2\times 1/4$ as chemical potential increases.
On the other hand, Hall resistance goes up as magnetic field $B$ increases. ) 
Therefore, when $T\neq 0$, the transition between plateaus in the variable $B$ is not sharp as shown in Fig.\ref{b}.

In the figure, we show the width $\Delta B$ between two plateaus. We expect from the above argument that the width $\Delta B \to 0$, as $T \to 0$.
Experiments\cite{deltaB} have shown that the width $\Delta B \propto T^{\kappa} \to 0$ as $T \to 0$ with $\kappa \sim 0.4$
as far as $T$ is not small so much. However,
experiments\cite{sat1,sat2,sat3,sat4,sat5} in detail indicate that $\Delta B$ approaches non zero constant as $T \to 0$.
That is, the width saturates at low temperature.
There is not yet convincing reason why the width saturates at finite value without vanishing. 
The similar phenomena can be seen as a decoherence\cite{decoherence} in states of electrons in topological insulators, e.c.t..
It is considered that the saturation arises due to intrinsic origin, not extrinsic one: Effect of temperature induced in measurement, finite size effect of Hall bar, 
e.c.t. have been excluded from the cause of the saturation. But, there is no plausible explanation of the intrinsic origin.
The presence of the finite width $\Delta B$ even at $T=0$ shows the presence of a decoherence mechanism.
Especially, it indicates non vanishing probability of
a transition from an electron in localized state
to an electron in extended state. Such a transition may be caused by axion dark matter.

The plateau-plateau transition is a kind of phase transition. Actually, it has been shown\cite{aoki1,aoki2} that
a coherent length $\xi(E)$ diverges such as $\xi(E)\to |E-E_{n\pm}|^{-\nu}$
as $E\to E_{n\pm}$ with $\nu\sim 2.4$. The coherent length may be considered to be typical length of localized state
with energy $E$.
The transition is characterized as the one between insulator and metal. The state with $E_f<E_{1+}$ is insulator, while the state with   
$E_f \ge E_{1+}$ in which electrons occupy the states with $E=E_{1+}$, is metal. 
The divergence of the coherent length at $E=E_{n\pm}$ implies that 
only a single state with energy $E_{n\pm}$ is the extended state, in which electrons can carries electric current in Hall bar with infinite length.

\vspace{0.1cm}
The coherent length is a typical size of localized state with energy $E$. Thus,
the size of localized state becomes larger than the finite size of Hall bar as $E$ approaches critical energies $E_{n\pm}$. 
It implies that localized states with their energies $E$
in the range $E_{n\pm}-\delta \le E \le E_{n\pm}+\delta$ can be effective extended states, 
which carry electric currents from boundary to boundary in Hall bar with finite size.
Then, insulator metal transition arises when Fermi energy $E_f$ approaches the energy $E_{n\pm}-\delta$ from below or above. 
Without thermal effect, the transition is expected to be sharp. But it is not true in
actual Hall bars. The width $\Delta B$ saturates at finite value as $T\to 0$. Finite size of Hall bar\cite{sat3} causes the saturation, but
the finite size effect is not
only the cause\cite{sat4,sat5}.
In this paper we discuss that the axion dark matter is the cause of the saturation.

\vspace{0.1cm}
Axion dark matter is converted to radiations in the presence of magnetic field. In particular, 
Such radiations are inevitably present in the experiment of
quantum Hall effect, because strong magnetic field generated by superconducting magnet is used. 
The magnet plays a role of cavity, with use of which axion dark matter has been explored.
In this paper we consider QCD axion, whose mass $m_a$ is supposed to be in the range $10^{-4}\rm eV \sim 10^{-6} eV$.
Among them, we specially focus on the mass range  $10^{-5}\rm eV \sim 4\times10^{-6} eV$ because
previous experiments about the width suggests the presence of the axion mass in the range, see below.
( The frequencies of corresponding radiations are 
in about $1\mbox{GHz} \sim 2.4\mbox{GHz}$. They are microwaves. ) 
Radiation energies converted from axions are almost the same as $m_a$, because the momenta
of the dark matter axions are of the order of $10^{-3}m_a$.
Thus, electrons with energy $E$ acquires the energy $E+m_a$ when they absorb the radiations.

\vspace{0.1cm}
We make a comment. 
Two dimensional electrons are formed in a quantum well fabricated by semiconductor. The width of the well 
is less than of the order of $10^{-6}$cm. The semiconductor
is transparent for microwaves in low temperature.

\vspace{0.1cm}
We consider the case in which
Fermi energy is slightly smaller than $E_{1+}-\delta$.
By absorbing the radiations from axions,
electrons with energies $E$ ( $<E_f-\delta$ ) transit to the states with the energies $E+m_a$ larger than $E_{1+}-\delta$.
The process can arise only for $m_a+E_f >E_{1+}-\delta$. 
The transited electrons with energies in the range $E_{n\pm}-\delta \le E \le E_{n\pm}+\delta$ carry electric current. 
There are non vanishing probability that
the effective extended states are occupied, 
even if the Fermi energy $E_f$ less than $E_{1+}-\delta$.  
Thus, 
the axions may change the value of Hall resistance. The effect is similar to thermal one with temperature $T\sim m_a$.

\vspace{0.1cm}
Here we should stress that even if only a single electron occupies an effective extended state,
Hall resistance or conductance changes dramatically to show plateau-plateau transition\cite{topology1,topology2}.
Although the probability of axion conversion to photon is extremely small, the possibility of a single electron 
making transition to an effective extended states
is not small because there are a number of localized electrons 
which may transit to the effective extended states with energies $E+m_a$ ( $E_{1+}-\delta <E+m_a<E_{1+}+\delta$ )
in Hall bar.

%

\section{axion dark matter}
\label{5}
In order to discuss the electron transition by absorbing radiations from axion dark matter, we explain
electromagnetic coupling with axion\cite{axion1,axion2,axion3} and smallness of the coupling.  
The coupling $g_{a\gamma\gamma}$ between axion $a(t,\vec{x})$ and electromagnetic field is 
given by

\begin{equation}
\label{La}
L_{a\gamma\gamma}=g_{a\gamma\gamma}a(t,\vec{x})\vec{E}\cdot\vec{B},
\end{equation}
with electric $\vec{E}$ and magnetic $\vec{B}$ fields.
We note that $g_{a\gamma\gamma}=g_{\gamma}\alpha/f_a\pi$,
where $\alpha\simeq 1/137$ denotes fine structure constant and $f_a$ does axion decay constant
satisfying the relation $m_af_a\simeq 6\times 10^{-6}\rm eV\times 10^{12}$GeV in the QCD axion.
The parameter $g_{\gamma}$ depends on the axion model, i.e. 
$g_{\gamma}\simeq 0.37$ for DFSZ model\cite{dfsz,dfsz1} and $g_{\gamma}\simeq -0.96$ for KSVZ model\cite{ksvz,ksvz1}.

The interaction term is much small in the case of the axion dark matter $a(t,\vec{x})\simeq a_0\cos(m_at)$. 
( The momentum $p$ of the axion dark matter is of the order of $10^{-3}m_a$ so that we may neglect the momentum. )
We estimate it supposing that the dark matter in the Universe is composed of the axion. The local energy density $\rho_d$ of the dark matter
is given such that
$\rho_d=m_a^2\overline{a(t,\vec{x})^2}=m_a^2a_0^2/2\sim 0.3\rm GeV/cm^3$ ( $\overline{Q}$ denotes
time average of the quantity $Q$. ) Thus, the coupling of the CP violating term is 
found to be much small; $g_{a\gamma\gamma}a(t,\vec{x})\sim 10^{-21}$
independent of the QCD axion mass. 
Thus, the electric field $E_a$ of the 
radiation produced\cite{sikivie, iwazaki01} in vacuum by the axion under external magnetic field $B$ is much weak such as $E_a\sim g_{a\gamma\gamma}aB$.
Such radiations inevitably arise in experiment of quantum Hall effect because strong magnetic field $ \sim 10$T is used.
Furthermore, superconducting magnet plays a role of cavity\cite{sikivie,iwazaki01} for the radiations.

As we have mentioned, we focus on the mass $m_a$ such as  $10^{-5}\mbox{eV} >m_a> 4\times10^{-6} $eV
( the corresponding frequency $m_a/2\pi$ of radiation is about $1\mbox{GHz} \sim 2.4\mbox{GHz}$. )

In the next section, we estimate transition probability of electron in a state with energy $E$ ( $<E_f$ ) to a state with energy larger 
than $E_{1+}-\delta$
by absorbing the radiation, $\vec{A}_a=\vec{A}_0\exp(-i\omega t+i\vec{p}\cdot\vec{x})\simeq \vec{A}_0\exp(-im_a t)$
whose amplitude $|\vec{A}_0|\sim g_{a\gamma\gamma}a_0B/m_a$. We take average over the direction of the gauge field $\vec{A}_0$
because the polarization of radiations produced in the experiment are not well controlled 
contrary to the case of the detection of axions in resonant cavity. The point is that
the magnitudes $|\vec{A}_0|$ of the radiations are extremely small.

\section{transition by absorption of electromagnetic wave caused by axion}
\label{6}

We discuss a transition from an electron in a state with energy $E_{\alpha}$ to a state with energy $E_{\alpha}+m_a$ ( $>E_{1+}-\delta$ ).
A state in a Landau level $n$ is described by the wave function in the following,

\begin{equation}
\Phi_{\alpha}=\exp(-iE_{\alpha}t)\int dk D_nf_{\alpha}(k)H_n\Big(\frac{x-l_B^2k}{l_B}\Big)\exp\Big(-\frac{(x-l_B^2k)^2}{2l_B^2}\Big)\exp(-iky),
\end{equation}
with $D_n=(2^{n+1}n!\pi^{3/2}l_B)^{-1/2}$.
The function $f_{\alpha}(k)$ is taken such that the wave functions are eigenstates of electron's Hamiltonian with potential $V$ representing disorder.
Normalization is such that that $1=\int dxdy \overline{\Phi_{\alpha}(x,y)}\Phi_{\alpha}(x,y)=\int dk \overline {f_{\alpha}(k)}f_{\alpha}(k)$.
The state is characterized by index $\alpha$. Here we assume no mixing between different Landau levels and between different spin states.
Electron in the state $\alpha$ makes a transition to a state with energy $E_{\alpha}+m_a$ by absorbing radiation. 
Here we only consider the Landau level $n=1$.
 
The absorption is described by
the following interaction Hamiltonian, 

\begin{equation}
\label{eq4}
H_a=\frac{-ie\vec{A}_a\cdot\vec{P}}{m^{\ast}}
\end{equation}
with $\vec{P}=-i\vec{\partial}+e\vec{A}_{ex}$ and $\vec{B}=\vec{\partial}\times \vec{A}_{ex}$,
where the gauge potential $\vec{A}_a$ is taken such that $\vec{A}_a\sim \vec{A}_0\exp(-im_at)$ with $|\vec{A}_0|\simeq g_{a\gamma\gamma}a_0B/m_a$.

\vspace{0.1cm}
As we can see soon later, the radiations with the electric fields $\propto \partial_0\vec{A}_a$ 
parallel to the magnetic field $\vec{B}$ are not absorbed in the Hall bar, while
radiations with the electric fields perpendicular to $\vec{B}$, that is, parallel to Hall bar, are absorbed.  
Thus, we have no factor in eq(\ref{eq4}) of permittivity of semiconductors.
Radiations produced by axions are not well controlled in experiments of quantum Hall effect, 
compared with those in resonant cavity experiment for searching axion.
There are several components of metals in dilution refrigerator used in the experiments. 
Thus, their electric fields are not necessarily parallel to magnetic field imposed on the Hall bar. 
Relevant radiations, which are absorbed in the Hall bar, are those with electric fields parallel to the Hall bar.

\vspace{0.1cm}
The transition amplitude from the state $\alpha$ to the state $\beta$ is proportional to 

\begin{equation}
<\beta |H_a|\alpha>=\int dxdy \overline{\Phi}_{\beta}(x,y)\frac{-ie\vec{A}_a\cdot\vec{P}}{m}\Phi_{\alpha}(x,y)=
i(E_{\beta}-E_{\alpha})e\vec{A}_a\cdot <\beta|\vec{x}|\alpha>\equiv i(E_{\beta}-E_{\alpha})e\vec{A}_a\cdot \vec{L}_{\alpha\beta}
\end{equation}
where $<\beta|z|\alpha>=0$ because the two dimensional states $\alpha$ and $\beta$ are orthogonal to each other. 
Electrons trapped in quantum well have wave function such as $D(z)\Phi_{\alpha}(x,y)$; $D(z)$ is the common factor for all $\alpha$.  

 $\vec{L}_{\alpha\beta}$( $\equiv <\beta|\vec{x}|\alpha>$ ) denotes a length scale of overlapping region of two states $\Phi_{\alpha}$ and $\Phi_{\beta}$,
which is defined in the following,

\begin{equation}
\vec{L}_{\alpha\beta}\equiv \int dxdy \overline{\Phi}_{\beta}(x,y)\vec{x}\Phi_{\alpha}(x,y).
\end{equation}

Consequently,
the number of electrons $N$ making transitions per unit time by absorbing the radiation $\vec{A}_a$
from states with energies $E_{\alpha}$ lower than Fermi
energy $E_f$ ( $<E_{1+}-\delta$ ) to states with energies $E_{\beta}$ larger than $E_{1+}-\delta$ is given by 

\begin{eqnarray}
\dot{N}&=&2\pi S^2\int dE_{\beta}dE_{\alpha}\rho({E_{\beta}})\rho(E_{\alpha})|<\beta |H_a|\alpha>|^2\delta(E_{\alpha}+m_a-E_{\beta}) \nonumber \\
&=&2\pi S^2\int dE_{\alpha}\rho({E_{\alpha}+m_a})\rho(E_{\alpha})m_a^2\Big(e\vec{A}_a\cdot \vec{L}_{\alpha\beta}\Big)^2
\end{eqnarray}
with surface area $S$ of the Hall bar. We take $S=500\mu \rm m\times 200\mu m=10^{-3}cm^2$ as typical value.


\vspace{0.1cm}
As we have explained, even localized states can carry electric current 
when the length scale $\xi(E_{\beta})$ of the state $\beta$
is larger than the length $L_h$ of Hall bar.
The transition of electron from the state $\alpha$ to the state $\beta$ increases Hall conductivity.
It has been discussed that the scale $\xi(E)$ behaves such as $\xi(E) \propto |E-E_{1+}|^{-\nu}$ 
as $E\to E_{1+}$ with $\nu \simeq 2.4$.  For instance,
it has been numerically shown\cite{aoki1,aoki2} using a finite size scaling analysis that $\xi(E)>100\mu $m for $B=10$T  
when $|E-E_{1+}|<0.1 \Delta E$ where $\Delta E$ denotes
the width of number density of states $\rho(E)$. For instance $\rho(E) \propto \sqrt{1-(E-E_{1+})^2/\Delta E^2}$.
Hence, even if the Fermi energy is less than $E_{1+}-\delta$, 
Hall conductivity $\sigma_{xy}=3\times e^2/2\pi$ jumps to  $\sigma_{xy}=4\times e^2/2\pi$,
if there are electrons occupying in the states $\beta$.
 
We consider the transition of electrons $\alpha$ with the energy $E_{\alpha}$ ( $<E_f-\delta$ ) to the states $\beta$ 
with the energy $E_{\beta}$ in the range $E_{1+}-\delta<E_{\beta}<E_{1+}+\delta$, assuming tentatively $\delta = \Delta E/10$,
in the analysis of the Hall bar with $L_h\sim 500\mu$m. 
We call the range as E range. Effective extended states with energies in the E range can carry electric current.
We should note that the width $\delta$ of the E range
depends on each Hall bar. In the present paper we assume $\delta \equiv \Delta E/10$.


We stress that even if only a single electron is present in the effective extended states, 
Hall resistance takes a value of
$\rho_{xy}=2\pi/e^2\times 1/4$. The probability that the states in the E range are occupied is non vanishing even if the Fermi energy 
is smaller than $E_{1+}-\delta$. Thermal fluctuation or electromagnetic noise by axions mentioned above make the probability non vanishing.
It leads to the finite width of $\Delta B$ in plateau-plateau transition even at $T=0$.

\section{number of electrons steady present in effective extended states}
\label{7}
Now, we estimate the transition probability of localized electrons to effective extended states. In particular,
we estimate the number $\dot{N}$ of electrons per unit time occupying the states in the E range,
$E_{1+}-\delta$ to $E_{1+}+\delta$. In the estimation we need to know the 
number density of the states $\rho(E)$. 
It is shown schematically in Fig.\ref{a}. The detail of the distribution
depends on each of Hall bars used in experiments. It is characterized by the width $\Delta E$ of the distribution.
Hereafter, we suppose the distribution

\begin{equation}
\rho(E)=\rho_0\sqrt{1-\Big(\frac{E-E_{1+}}{\Delta E}\Big)^2} \quad \mbox{with} \quad |E-E_{1+}|\le \Delta E 
\quad \mbox{otherwise} \quad \rho(E)=0
\end{equation} 
with $\rho_0=(eB/2\pi)\times 2/(\pi \Delta E)$,
where $\int dE \rho(E)=eB/2\pi$ represents the number density of electrons in a Landau level;
$\int dE \rho(E)=eB/2\pi\simeq 2.4\times 10^{11}/\rm cm^2$$(B/10T)$. The width $\Delta E$ \cite{width} is of the order of 
$\sim 10^{-4}$eV. Tentatively, we take $\delta=\Delta E/10\sim 10^{-5}$eV. Thus, the E range is much smaller than Landau level spacing 
$eB/m^{\ast}\simeq 1.7\times 10^{-2}\mbox{eV}(B/10\mbox{T})(0.067m_e/m^{\ast})$.

\vspace{0.1cm}

The length $\vec{L}_{\alpha\beta}$ denotes length scale of overlapping region between two localized states $\alpha$ and $\beta$.
It depends on both the states. In general, there is no overlapping when the energies $E_{\alpha}$ and $E_{\beta}$
are much far from the critical energy $E_{1+}$; $\vec{L}_{\alpha\beta}\simeq 0$. This is because the sizes of the states are small.  
The size of a localized state, however, becomes much larger as its energy approaches to the critical energy.
The size becomes infinity at the critical energy $E_{1+}$.
Because we consider the localized states $\alpha$ with their energies $E_{\alpha}$ ($<E_f$ ) such as $E_{\alpha}+m_a$ in the E range
( i.e. $E_{1+}-\delta-m_a <E_{\alpha}<E_{1+}+\delta-m_a $ ), the energies $E_{\alpha}$ are very close to the critical energy $E_{1+}$
because $m_a\sim \delta$ which we mainly consider as axion mass $m_a$ equal to or less than $10^{-5}$eV .  So,
their sizes are large and comparable to the effective extended states.
On the other hand, the states $\beta$ are in the E range so that their length scales are larger than 
the scale $L_h=500\mu$m$\sim 10^5l_B(10\mbox{T}/B)^{1/2}$ of the Hall bar.
Therefore, 
their overlapping length $L_{\alpha\beta}$ can be also large such as for instance $L_{\alpha\beta}\sim 10^3l_B$.  
For simplicity, we assume that the length $L_{\alpha\beta}$ is independent of the states $\alpha$ and $\beta$;
$L_{\alpha\beta}=A\l_B$ with $A=10^4$. 
Then, it follows that

\begin{equation}
\label{N}
\dot{N}=\int dE_{\alpha}2\pi S^2\rho({E_{\alpha}+m_a})\rho(E_{\alpha})m_a^2
\Big(e\vec{A}_a\cdot \vec{L}_{\alpha\beta}\Big)^2
\simeq 2\pi S^2m_a^2\Big(\frac{eB}{2\pi}\frac{2}{\pi \Delta E}\Big)^2e^2A_0^2A^2l_B^2
\times \Delta,
\end{equation}
where $\int dE_{\alpha}\equiv \Delta$ and
we put  $(e\vec{A}_a\cdot \vec{L}_{\alpha\beta})^2\equiv e^2A_0^2A^2l_B^2$ with $A_0=g_{a\gamma\gamma}a_0B/m_a$.

It is obvious that the integration $\int dE_{\alpha}$ is taken over the range 
$E_f \ge E_{\alpha}\ge E_{1+}-\delta-m_a $. Thus, $\Delta=E_f-(E_{1+}-\delta-m_a)$
for $E_{1+}+\delta -m_a \ge E_f$, otherwise $\Delta=2\delta$.

\vspace{0.1cm}
When the Fermi energy is smaller than $E_{1+}-\delta-m_a$, i.e. $E_f<E_{1+}-\delta-m_a$, 
there are no electrons transiting to states in the E range, i.e. $\dot{N}=0$ because $\Delta=0$. 
Hall conductance $\sigma_{xy}$ keeps the value $3e^2/2\pi$.
Once the Fermi energy goes beyond the value $E_{1+}-\delta-m_a$ ( $\Delta=E_f-(E_{1+}-\delta-m_a)>0$ ),
the probability that electrons occupy the effective extended states 
is non vanishing so that $\dot{N}>0$.
The Hall conductance $\sigma_{xy}$ gradually increases larger than $3e^2/2\pi$ as Fermi energy $E_f$ increases.
But the Fermi energy $E_f$ goes beyond $E_{1+}+\delta-m_a$ ( $E_f>E_{1+}+\delta-m_a$ ), 
$\dot{N}$ takes its maximum and does not increase,
keeping the value with $\Delta =2\delta$ in eq(\ref{N}). This is because electron in the state with energy $E_{\beta}=E_{\alpha}+m_a$
does not carry electric current.
Furthermore, when $E_f$ goes beyond the value $E_{1+}-\delta$, $\sigma_{xy}$ becomes  $4e^2/2\pi$
because there are electrons always present in the effective extended states without the transition.
Then, the plateau $\sigma_{xy}=4\times e^2/2\pi$ newly arises. 
The width of Fermi energy $\Delta E_f=E_{1+}-\delta-(E_{1+}-\delta-m_a)$ is given by
$\Delta E_f=m_a$. Within the width, plateau-plateau transition takes place. It corresponds to the width $\Delta B$.
Later we present the explicit relation $\Delta E_f \propto \Delta B$ between $\Delta E_f$ and $\Delta B$.


Typically, the value of $\dot{N}$ is 

\begin{equation}
\dot{N}\simeq 3\times 10^4\, \mbox{s$^{-1}$}\Big(\frac{A}{10^4}\Big)^2 \Big(\frac{S}{\mbox{10$^{-3}$cm$^2$}}\Big)^2
\Big(\frac{10^{-4}\mbox{eV}}{\Delta E}\Big)^2
\Big(\frac{\rho_d}{0.3\,\mbox{GeVcm$^{-3}$}}\Big)\Big(\frac{B}{10\rm T}\Big)^3\Big(\frac{\delta}{10^{-5}\rm eV}\Big)
\end{equation}
where we simply assumed $\Delta=2\delta$.
The value has been obtained under the assumption that the size of Hall bar $L_h\simeq 500\mu$m 
and the length scale, tentatively, $Al_B\sim 10^4l_B$ of the overlapping 
between localized states $\alpha$ and effective extended states. The energies $E_{\alpha}$ 
of the localized states $\alpha$ is very near the E range  ( $E_{1+}-\delta \sim E_{1+}+\delta$ ) because $E_{\alpha}+m_a$ must be
in the E range.  Being close is necessary for the overlapping $L_{\alpha\beta}=Al_B$ to be large such as $A=10^4$.
As we can see, it is unclear how large the parameter $A$ is.
However, it turns out later that the assumption of large $A$ is not irrelevant for the estimation of
the number of electrons steady occupying the effective extended states. 
Electrons steady occupying the states are those defined such that they are determined by the balance between incoming electrons
to the states and outgoing electrons from the states. 

\vspace{0.1cm}
( In the above estimation, we have assumed that the radiations $\vec{A}_a$ absorbed by electrons are generated by axion dark matter.
When the radiations are externally imposed on Hall bar by ourselves, we find that the formula $\dot{N}$ is proportional to $\nu^2$ with the frequency $\nu$ of
the radiations. )

\vspace{0.1cm}
We have estimated the transition probability of localized electrons to effective extended states by absorbing radiations from axion dark matter
at the temperature $T=0$.
But, such transition also arises owing to the black body radiation when $T\neq 0$.
 The energy density of the radiations are much bigger than that of
radiations from axions. The effect of black body radiation may be considered as thermal noise.
Thus, we should take into account the thermal noise.

\vspace{0.1cm}
The energies $\omega$ of the black body radiations are approximately restricted to be smaller than the temperature $T$; $\omega<T$, because
the number of the radiations with energies much larger than $T$ are exponentially suppressed. For simplicity, we only consider
the black body radiations with energies $\omega <T$. 
Localized electrons with energies $E_{\alpha}$ less than Fermi energy $E_f$ can be transited to effective extended states by absorbing the
radiations only when $E_{\alpha}+\omega>E_{1+}-\delta$. These electrons contribute Hall conductance.
Thus, when the temperature is sufficiently large such as $T>E_{1+}-\delta -E_f$, 
the radiations with the energies $\omega$ in the range, $T>\omega>E_{1+}-\delta -E_f$, are absorbed 
and the Hall conductance increases. The energy power $P_n$ of the noise is given by $P_n=T(T-(E_{1+}-\delta -E_f))$.
On the other hand, when the temperature is less than $E_{1+}-\delta -E_f$, the black body radiations do not contribute
the increase of the Hall conductance. ( Even if localized electrons absorb the radiations, they are only transmitted to localized states, not to effective 
extended states. )
Therefore, the increase of the Hall conductance is only caused by the dark matter axion when temperature $T$ is lower than
$E_{1+}-\delta -E_f$ and the axion mass $m_a$ is larger than $E_{1+}-\delta -E_f$, i.e. $m_a>E_{1+}-\delta -E_f>T$. 
When the temperature increases larger than $E_{1+}-\delta -E_f$, black body radiations also cause the increase of the Hall conductance.
Even if the Fermi energy $E_f$ is very near to but smaller than 
$E_{1+}-\delta$, plateau-plateau transition does not take place at $T=0$, while when $E_f \ge E_{1+}-\delta$, the transition takes place.
So, without axion effect, the feature of the transition is like step function.  
The axion effect makes the transition smooth so that the width $\Delta B$ of the transition is non vanishing even at $T=0$.

\vspace{0.1cm}
Obviously, the axion effect dominates the thermal effect for sufficiently low temperature, while
the thermal effect dominates the axion effect for high temperature. The width $\Delta B(T)$ is determined by
the dominant effect. Actually, the width decreases as the temperature decreases when it is very high. It implies that the thermal effect
controls the width. But the width saturates at a critical temperature below which the width does not decreases more.
The axion effect dominates over the thermal effect below the critical temperature. 
If the thermal effect dominates over the axion effect even beyond the critical temperature, the width $\Delta B(T)$
decreases and does not saturate. 

We have the energy power generated by the dark matter axion such as
$P_a=m_a\dot{N}$ and thermal noise $P_T=T(T-(E_{1+}-\delta-E_f))$. The axion effect with the power $P_a$ 
makes localized electrons transit to
effective extended states, while thermal effect with power $P_T$ also makes localized electrons transit to the extended states. 

Here,
we would like to know the critical temperature at which the effect of the temperature and the effect of axion are balanced in actual measurement.
It should be stressed that each measurement of Hall conductivity for $B$ and $T$ fixed is performed with several seconds.
Within the period, the signal of the axion should be compared with the thermal noise. 
( Even if the signal is much weak, it is observed when sufficiently long period is taken for the measurement. )  
We suppose that the Fermi energy $E_f\simeq E_{1+}-\delta $ by appropriately adjusting magnetic field $B$. 
Then,
we calculate the ratio of the signal to the noise. 

The energy power generated by the dark matter axion in the Hall bar is 

\begin{equation}
P_a\equiv \dot{N}m_a\sim 4.8\times 10^{-20}\mbox{W}\Big(\frac{A}{10^4}\Big)^2 \Big(\frac{S}{10^{-3}\mbox{cm$^2$}}\Big)^2
\Big(\frac{10^{-4}\mbox{eV}}{\Delta E}\Big)^2
\Big(\frac{\rho_d}{0.3\,\mbox{GeVcm$^{-3}$}}\Big)\Big(\frac{B}{10\rm T}\Big)^3\Big(\frac{\delta}{10^{-5}\rm eV}\Big)\Big(\frac{m_a}{10^{-5}\mbox{eV}}\Big)
\end{equation} 

\vspace{0.1cm}
We compare it with thermal noise, $P_T=T(T-(E_{1+}-\delta-E_f))$. 
Assuming $T \gg E_{1+}-\delta-E_f$,  
SN ratio is given with $T=10$mK ( $\simeq 8.6\times 10^{-7}$eV ),

\begin{eqnarray}
&&\frac{P_a\sqrt{(T-(E_{1+}-\delta-E_f) \times 1\mbox{s}}}{P_T} \nonumber \\
&\sim& \frac{P_a\sqrt{T \times 1\mbox{s}}}{T^2}
\simeq 	9.8\times \Big(\frac{A}{10^4}\Big)^2\Big(\frac{S}{10^{-3}\mbox{cm$^2$}}\Big)^2\Big(\frac{10^{-4}\mbox{eV}}{\Delta E}\Big)^2\Big(\frac{\rho_d}{0.3\,\mbox{GeVcm$^{-3}$}}\Big)\Big(\frac{B}{10\rm T}\Big)^3\Big(\frac{\delta}{10^{-5}\rm eV}\Big)\Big(\frac{m_a}{10^{-5}\rm eV}\Big) 
\end{eqnarray}

Thus, we find that the axion effect can be observable at the low temperature $\sim 10$mK or less. The saturation of the width $\Delta B$ arises
around temperature $\sim10$mK. Although the ratio $S/N$ depends on the parameters of Hall bar, we expect that 
the saturation can be seen at the low temperature for relatively large Hall bar such as $500\mu\mbox{m}\times 200\mu\mbox{m}$.
Actually, a measurement in detail of the saturation has been performed\cite{sat6} using Hall bars with similar sizes mentioned here. It  has shown that
the saturation of the width $\Delta B$ is observed at low temperature $\sim 30$mK. It has also been shown that
the cause of the saturation is not finite size effect of Hall bar. The experiment is consistent with our argument that
the saturation at the low temperature is caused by axion dark matter.

( Our estimation of the SN ratio is similar to the estimation in cavity experiment for axion detection. The experiment of quantum Hall effect
measures transverse electric currents $I_{xy} $ or voltage $V_{xy}$ to obtain Hall conductivity, while the cavity experiment
also measure electric current or voltage to obtain the power generated by axion. Therefore, the estimation of the SN ratio
should be identical to each other. )



\vspace{0.1cm}
We have obtained the number of electrons transited from localized states $\alpha$ to effective extended states $\beta$.
These electrons do not remain to stay in the effective extended states. They loose their energies and go back to localized states.
It is difficult to find how fast they loose their energies because there are several mechanisms of the energy loss;
energy loss by emission of electromagnetic radiation, emission of phonon, or by interaction with other electrons, e.t.c..

\vspace{0.1cm} 
Here, we only consider a mechanism of energy loss by emission of electromagnetic radiations such that 
the electrons in the states $\beta$ go back to the original localized states $\alpha$ 
by spontaneously emitting radiations. Such a transition is possible because the localized states $\alpha$ are unoccupied.
The mechanism of the energy loss can be explicitly treated. 
We would like to find the average number of electrons staying in the effective extended states.
The number is determined by the balance between incoming electrons and outgoing electrons.
Because we have obtained the incoming electrons $\dot{N}$, we need to know the number of outgoing electrons.

 So, we calculate the probability that an electron in the state $\beta$ with energy $E_{\beta}$ 
( $E_{1+}+\delta \le E_{\beta} \le E_{1+}-\delta $ ) transits to the state $\alpha$ with energy $E_{\alpha}=E_{\beta}-m_a$ by emitting
radiation with energy $m_a$.

That is, we calculate the transition amplitude,

\begin{equation}
\langle E_{\alpha}, k,\gamma |\frac{e\vec{A}\cdot \vec{P}}{m^{\ast}}|E_{\beta},vac \rangle=i(E_{\alpha}-E_{\beta})\langle E_{\alpha}|\vec{x}|E_{\beta} \rangle 
\cdot \langle k, \gamma |e\vec{A}|vac \rangle
\end{equation}
where the state $|k, \gamma \rangle $ denotes
a photon with momentum $k=|\vec{k}|$ and polarization $\gamma$. The radiation is described by the gauge potential $\vec{A}$.
We make approximation $\exp(i\vec{k}\cdot \vec{x})\simeq 1$ because
wave length $k^{-1}=m_a^{-1}\simeq 2$cm $(10^{-5}$eV$/m_a)$ of the radiation is much larger than the scale of Hall bar.

Then,
by summing over all directions of momentum $\vec{k}$ with $|\vec{k}|=E_{\beta}-E_{\alpha}$ and polarization $\gamma$, 
we find that the transition probability per unit time $\dot{T_p}$ is

\begin{equation}
\dot{T_p}=\frac{4\alpha}{3}|\vec{L}_{\alpha\beta}|^2m_a^3=\frac{4\alpha}{3}A^2l_B^2m_a^3
\end{equation}
where we have used the length scale $|\vec{L}_{\alpha\beta}|^2=A^2l_B^2$ of the overlapping region between the state $\beta$ 
and the state $\alpha$.
Thus, we obtain

\begin{equation}
\dot{T_p}\sim 2.2\times 10^3\, \mbox{s}^{-1}
\Big(\frac{m_a}{10^{-5}\mbox{eV}}\Big)^3\Big(\frac{10\mbox{T}}{B}\Big)\Big(\frac{A}{10^4}\Big)^2. 
\end{equation} 

By absorbing radiations with the energy $m_a$, localized electrons make transition to effective extended states, 
while the electrons go back to
the original localized states by spontaneously emitting radiations with the energy $m_a$. 
Consequently, a stationary state is realized. 
We have estimated both numbers of incoming and outgoing electrons 
in the effective extended states.
Average number $N_{av}$ of electrons stationary occupying the
effective extended states is determined by the valance between incoming electrons $\dot{N}$ 
and outgoing electrons $N_{av}\dot{T_p}$, i.e. $\dot{N}=N_{av}\dot{T_p}$,

\begin{equation}
\label{14}
N_{av}=\frac{\dot{N}}{\dot{T_p}}\sim 14 \Big(\frac{S}{\mbox{10$^{-3}$cm$^2$}}\Big)^2
\Big(\frac{10^{-4}\mbox{eV}}{\Delta E}\Big)^2
\Big(\frac{\rho_d}{0.3\,\mbox{GeVcm$^{-3}$}}\Big)\Big(\frac{B}{10\rm T}\Big)^4\Big(\frac{\delta}{10^{-5}\rm eV}\Big)
\Big(\frac{10^{-5}\mbox{eV}}{m_a}\Big)^3.
\end{equation} 

It is important to note that the final result $N_{av}$ does not depend on the size of overlapping region $Al_B$. 

\vspace{0.1cm}
Even if Fermi energy is less than $E_{1+}-\delta$, there is the presence of
electrons in the effective extended states.  It leads to the finite width $\Delta B$ in plateau-plateau transition. 
Although the value $N_{av}$ depends on magnetic field $B$, size of Hall bar, or disorder potential $V$ ( namely $\delta$ and $\Delta E$ ),
the rough estimation implies that the axion dark matter can be a cause of the saturation of the width $\Delta B$ in 
plateau-plateau transition. This is because
the value $N_{av}$ is of the order of $1$ and only if a single electron occupies an effective extended state, 
the Hall conductivity changes by large step.
Although radiations from axions are extremely weak,
a large number of electrons of the order of $S\times eB/2\pi\sim 10^8(B/10\mbox{T})(S/10^{-3}\rm cm^2)$
are involved in the transition. Thus, a small fraction of localized electrons transiting to effective extended states
may change the Hall conductivity by large step.

\vspace{0.1cm}
As we have explained, incoming electrons $\dot{N}$ gradually increases as the Fermi energy increases from $E_{1+}-\delta-m_a$.
Accordingly, the number $N_{av}$ of electrons increases,
because the transition probability $\dot{T_p}$ does not depend on the Fermi energy.
The Hall conductivity also increases as $N_{av}$ increases.
In this way, the finite width $\Delta B(E_f)$ arises in the plateau-plateau transition.

\vspace{0.1cm}
We make a comment that we have not yet clarified the explicit relation between $\Delta B$ and $N_{av}$. 
We simply state that when $N_{av}=0$, $\sigma_{xy}=3\times e^2/2\pi$, but as $N_{av}$ increases more than $0$, 
$\sigma_{xy}$ increases more than $3\times e^2/2\pi$. Finally, when $N_{av}$ reaches at $1$, Hall conductivity takes the value
$\sigma_{xy}=4\times e^2/2\pi$.
The increase of the Fermi energy $E_f$ makes $N_{av}$ increase. Thus, the width $\Delta E_f$ corresponds to
the range of the value $N_{av}$ from $0$ to $1$. It also corresponds to the width $\Delta B$.
Later we derive the explicit relation between $\Delta E_f$ and $\Delta B$. 
But we do not know how $N_{av}$ depends on the Fermi energy $E_f$.
In other words, when the value $N_{av}$ changes from $0$ to $1$, what amount of the value $E_f$ changes, i.e. $\Delta E_f$.

\vspace{0.1cm}
The above estimation of $\dot{N}_{av}$ is very naive one. Electrons in the effective extended states $\beta$ loose their energies not only by transition to the
original localized states $\alpha$,
but also by transition to other localized states with energy larger than $E_{\alpha}$ ( $E_{\beta}=E_{\alpha}+m_a$ ). 
That is, they loose their energies by emissions of radiations or phonons with energies less than $m_a$. 
( In this case, $\dot{T}$ does not depend on $m_a$ ).
Furthermore, electrons loose their energies by colliding with other electrons,
transiting to localized states. The precise estimation of real transition probability $\dot{T}_{rp}$ is
difficult and beyond our scope. 

\vspace{0.1cm}
But we need to take into account the following fact in the estimation.
The transitions to localized states with energies less than Fermi energy $E_f$ are not allowed. This is because
the localized states are occupied. 
The transition is only allowed such that electrons in effective extended states transit to localized states with energies 
$E_{\alpha}$ such as $E_{1+}-\delta > E_{\alpha}>E_f$.
The states are not occupied. Of course, the transition to the effective extended states with energies
larger than $E_{1+}-\delta$ is also allowed. But the transition does not change the number of electrons steady staying in
the effective extended states. 
With these consideration,
we need to examine how the allowed real transition probability $\dot{T}_{rp}$ to localized states depends on the Fermi energy $E_f$,
in order to see the dependence of $N_{av}=\dot{N}/\dot{T}_{rp}$ on $E_f$.


\vspace{0.1cm}
The dependence of $N_{av}$ on $E_f$ arises not only from $\dot{N}$, but also $\dot{T}_{rp}$. 
Then, when the Fermi energy $E_f$ begins to go beyond the energy $E_{1+}-\delta-m_a$, $\dot{N}$ gradually increases more than $0$, but
$\dot{T}_{rp}$ is still
large enough for $N_{rav}=\dot{N}/\dot{T}_{rp} \ll 1$. This is because the range $E_{1+}-\delta-E_f$ is large.
$\dot{N}$ increases more but $\dot{T}_{rp}$ decreases as $E_f$ increases more beyond $E_{1+}-\delta-m_a$.
The reason of the decrease of $\dot{T}_{rp}$ is that
the number of localized states with energies $E$ within the range $E_{1+}-\delta > E>E_f$, decreases as $E_f$ increases.
Finally, $N_{rav}$ approaches its maximum when $E_f$ approaches $E_{1+}-\delta$. At the point, $\dot{T}_{rp}=\dot{T}$
because the range $E_{1+}-\delta-E_f$ vanishes.
The point corresponds to the end of plateau-plateau transition.
Therefore, actual relation between $\Delta E_f$ and $m_a$ is more complicated than the naive one 
$\Delta E_f=m_a$ mentioned above.

\vspace{0.1cm}
We may infer a real relation from a microwave experiment\cite{micro}, in which microwaves with frequencies $\nu$ are imposed on Hall bar. 
It has been observed that
the relation $\Delta B\propto \nu^{\gamma}$ with $\gamma \sim 0.4$. 
As we will demonstrate soon below, 
the relation $\Delta B\propto \Delta E_f$ holds in general. 
Thus, the experiment shows that $\Delta E_f\propto \nu^{\gamma}$ with $\gamma \sim 0.4$.
The microwave may be identified as the one converted from
the axion dark matter. Hence, $\Delta E_f\propto m_a^{\gamma}$.
This is the observed relation between axion mass $m_a$ and the width of Fermi energy $\Delta E_f$. 

\vspace{0.1cm}
According to the above argument, we speculate that the width $\Delta E_f$ does not depend on Landau level $n$. It is determined only by the
difference between $E_{n\pm}-\delta$ and $E_f (>E_{n\pm}-\delta-m_a$ ). In next section, we present an observation which shows that 
it is true.

\section{relation between $\Delta E_f$ and $\Delta B$ and its dependence on Landau level}
\label{8r}

There has been an observation\cite{sat6} which shows how the width $\Delta B(n)$ depends on plateau-plateau transition characterized with 
$\rho_{xy}=(2\pi/e^2)\times 1/n$; the dependence of $\Delta B(n)$ on $n$, which represents the transition width 
from the plateau $\rho_{xy}=(2\pi/e^2)\times 1/n$
to the plateau $\rho_{xy}=(2\pi/e^2)\times 1/(n+1)$.
First, we show the relation between $\Delta B(n)$ and Fermi energy $\Delta E_f$. 
According to the observation, it turns out that $\Delta E_f/\Delta E$  
is independent on Landau level $n$. ( $\Delta $E is the width of the density of states
$\rho(E)$. ) 

\vspace{0.1cm}
For simplicity, we assume that Zeeman energy is negligibly small. Thus, there are the number $2eB/2\pi$ of states degenerates
in each Landau level with energy $E_n=\omega_c (n+1/2)$ with $n\ge 0$. The degeneracy is lifted up by potential $V$.
The density of states $\rho_n(E,B)$ is characterized by the bandwidth $\Delta E$. We assume the density
$\rho_n(E,B)=2\rho_0\sqrt{1-((E-E_n)/\Delta E)^2}$ as used above; the factor $2$ comes from spin degree of freedom. 

Suppose that Landau levels up to $n$ are occupied and that the half of the Landau level $n+1$ is also occupied
at the magnetic field $B=B_c$. That is, the Fermi energy is located at the center of the number density $\rho_{n+1}(E,B)$ 
of Landau level $n+1$ at $B=B_c(n)$; $E_f=E_{n+1}$. We assume that $\delta/\Delta E \ll 1$ and
we neglect $\delta$.
Now, we make magnetic field $B$ increase such as $B_c(n)+\Delta B$. 
Then, the number of electrons occupying each Landau level $m ( \le n )$ increases 
so that the Fermi energy $E_f$ decreases such as $E_f=E_{n+1}-\Delta E_f$. Then,
by noting that the total number density $\rho_e$ of electrons does not change, it follows that

\begin{eqnarray}
\frac{2eB_c}{2\pi}(n+\frac{1}{2})&=&\frac{2e(B_c(n)+\Delta B)}{2\pi}n+\int_{E_{n+1}-\Delta E}^{E_{n+1}-\Delta E_f} dE \rho_{n+1}(E,B_c(n)+\Delta B) \\
&\simeq &\frac{2e(B_c(n)+\Delta B)}{2\pi}n+\frac{1}{2}\frac{2e(B_c(n)+\Delta B)}{2\pi}-2\rho_0\Delta E_f
\end{eqnarray}
with $\int_{E_{n+1}-\Delta E_f}^{E_{n+1}} dE \rho(E,B_c(n)+\Delta B)\simeq 2\rho_0 \Delta E_f$
for $\Delta E_f/\Delta E \ll 1$.

 It turns out that
 
\begin{equation}
\label{delta}
\Delta B(n)=\frac{4B_c(n)}{(2n+1)\pi}\frac{\Delta E_f}{\Delta E}.
\end{equation} 
 
It describes how the width $\Delta B(n)$ depends on Landau level $n$ and the width of Fermi energy $\Delta E_f$.
The relation is very general. Although we have used the explicit formula $\rho(E)\propto \sqrt{1-(E-E_n)^2/(\Delta E)^2}$,
we can derive similar result i.e. $\Delta B \propto \Delta E_f$, 
even if we use a different formula such as $\rho(E)\propto \exp(-(E-E_n)^2/(\Delta E)^2)$.

\vspace{0.1cm}
As we have mentioned, the width $\Delta E_f$ satisfies $\Delta E_f \propto m_a^{\gamma}$ with $\gamma\sim 0.4$. 
Setting $\Delta E_f=C\times m_a^{\gamma}$ with a constant $C$, we have 

\begin{equation}
\label{18}
\Delta B(n)=\frac{4B_c(n)}{(2n+1)\pi}\frac{Cm_a^{\gamma}}{\Delta E}.
\end{equation}

We find that the formula in eqs(\ref{delta}) and (\ref{18}) well agrees with the observation\cite{sat6}.
That is, the factor $\Delta E_f/\Delta E=Cm_a^{\gamma}/\Delta E$ is independent on $n$.
Here, we present the result $\Delta E_f/\Delta E\sim 0.27$ in the observation, 

\begin{eqnarray}
\frac{(2\times 2+1)\pi}{4}\frac{\Delta B(2)}{B_c(2)}&\simeq& \frac{5\pi}{4}\frac{0.7\mbox{T}}{11\mbox{T}}\simeq 0.25, \quad
\frac{(2\times 3+1)\pi}{4}\frac{\Delta B(3)}{B_c(3)}\simeq \frac{7\pi}{4}\frac{0.4\mbox{T}}{8\mbox{T}}\simeq 0.27  \nonumber \\
\frac{(2\times 4+1)\pi}{4}\frac{\Delta B(4)}{B_c(4)}&\simeq& \frac{9\pi}{4}\frac{0.23\mbox{T}}{6\mbox{T}}\simeq 0.27, \quad
\frac{(2\times 5+1)\pi}{4}\frac{\Delta B(5)}{B_c(5)}\simeq \frac{11\pi}{4}\frac{0.16\mbox{T}}{5\mbox{T}}\simeq 0.28 . 
\end{eqnarray}

\vspace{0.1cm}
Because the width $\Delta E$ \cite{andouemura1} in the density of state $\rho(E)$
does not depend on Landau levels $n$, $\Delta E_f$ does not depend on $n$ as we have discussed in the previous section.
The observation shows the
validity of our previous discussion.
 
Furthermore, the independence of  $\Delta E_f/\Delta E$ on $n$ shows that the width $\Delta B(n)$ becomes
smaller such as $\Delta B(n) \propto 1/n^2$ as $n$ increases, because $B_c(n)\propto n^{-1}$.
Plateaus in higher Landau levels are realized with smaller magnetic field $B$ when electron number density is constant.
Since the number $N_{av}$ in eq(\ref{14}) of electron steady present in extended states 
is proportional to $B^4$, the decrease of the width $\Delta B(n) \propto 1/n^2$ in higher Landau level is consistent with
our idea that the saturation of the width is caused by axion dark matter.  

\vspace{0.1cm}
Such a saturation of $\Delta B$ at low temperature have been observed in many experiments\cite{sat1,sat2,sat3,sat4,sat5} using various sizes
of Hall bar.
It is generally considered that the cause of the saturation is intrinsic. Similar phenomena, as is called phase decoherence, have been
observed\cite{decoherence} in various condensed matters. But there are no plausible explanations of the intrinsic cause.
In this paper we show that the saturation $\Delta B$ is caused by axion dark matter. It might be the intrinsic cause for
the phase decoherence observed in various condensed matters. ( Even without magnetic field, axion directly couples with electrons 
so that the electrons are in the bath of the axion dark matter at $T=0$. Hence, the coherence of electrons at $T=0$ may be disturbed. )

\section{determination of axion mass } 
\label{8}
 
Under strong magnetic field, axion is converted to electromagnetic radiation. Because experiment of quantum Hall effect uses such a strong magnetic
field, Hall bar is inevitably imposed by the radiation. We have discussed that the radiation produces 
finite width $\Delta B\neq 0$ at temperature $T=0$ in the plateau-plateau transition region.

\vspace{0.1cm}
The saturation of the width ( $\neq 0$ ) has been observed, contrary to the expectation that the width behaves
such as $\Delta B \propto T^{\kappa} \to 0$ for $T \to 0$ with $\kappa \sim 0.4$. 
How large the width is depends on each Hall bar. In order to suppress the finite size effect, 
we should use large Hall bar such as $L_h>500\mu$m.

Similar saturation has been observed
in a different situation. Namely,
when we impose microwaves on Hall bar, the width also depends on the frequency of the microwave\cite{micro}.
The width increases as the frequency increases, in other words, it decreases with the decrease of the frequency.
It has been shown at sufficiently low temperature that the width behaves such as 
$\Delta B \propto \nu^{\gamma}$ for $\nu=10\mbox{GHz} \sim 2\mbox{GHz}$ with $\gamma \sim 0.4$.
But, the width appears to be saturated at lower frequency $\nu <2$GHz.
Obviously, the frequency of the radiation plays a similar role of temperature.
The experiment is very suggestive when we notice 
that radiation from axion can be identified as radiation imposed externally on Hall bar.
Without radiations imposed externally, we still have radiations from axion dark matter in experiments of quantum Hall effect.

\vspace{0.1cm}
( We make a comment that the observational results of $\Delta B \propto \nu^{\gamma}$ with $\gamma \sim 0.4$ is consistent with
our estimation that $\dot{N}_{av}\propto \nu^2$ in the case of the radiations being externally imposed. 
The increase of the frequency makes $\dot{N}_{av}$ large so that $\Delta B$ becomes large. )

\vspace{0.1cm}
Our proposal for the determination of axion mass is in the following.
We impose microwave on Hall bar and measure Hall conductivity at sufficiently low temperature for
the width $\Delta B$ to be saturated without microwaves. The size of Hall bar should be larger than $500\mu$m.
That is, the experiment is performed when
the width $\Delta B$ has already been saturated at extremely low temperature. 
The temperature should be below several $10$mK because the effect of black body radiation
is less effective than the effect of the radiation produced by axion. 
We note that
the microwave \cite{micro} 
plays a role of temperature. That is, the width behaves such that $\Delta B \propto \nu^{\gamma}$ with $\gamma \sim 0.4$.

\vspace{0.1cm}

\begin{figure}[htp]
\centering
\includegraphics[width=0.6\hsize]{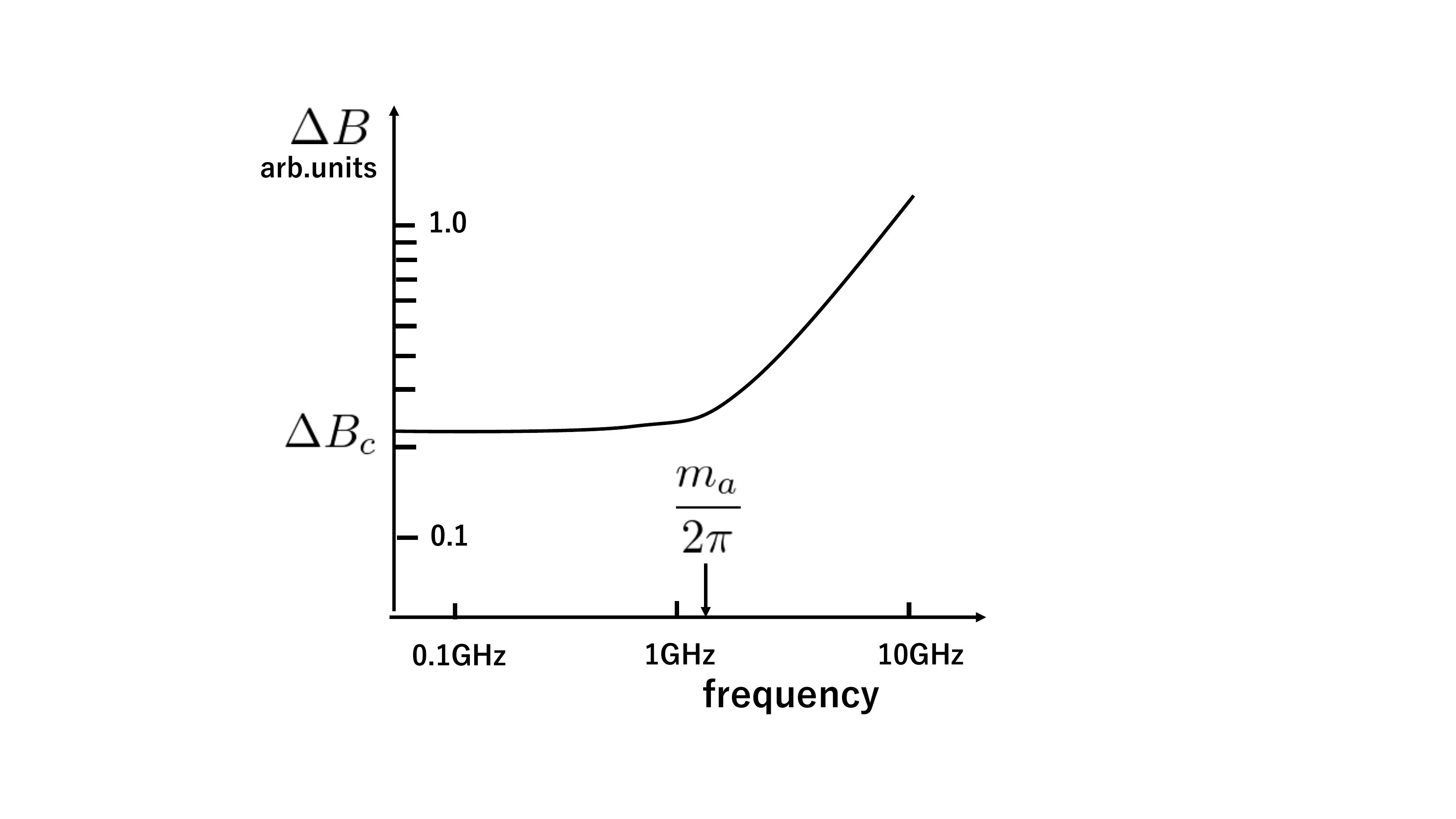}
\caption{saturation of $\Delta B$ in radiation frequency}
\label{c}
\end{figure}

The width $\Delta B$ decreases as the frequency of the microwave decreases as well as 
as the temperature decreases. 
But at sufficient low temperature, it saturates at a finite value $\Delta B_c$ in the absence of the external
microwaves.  On the other hand, at such a sufficient low temperature, the external microwave
leads to the width larger than $\Delta B_c$. The decrease of the frequency $\nu$ leads to the decrease of the width $\Delta B$
such that $\Delta B\propto \nu^{\gamma}$.
When the frequency is below a critical frequency, the width saturates at $\Delta B=\Delta B_c$ as shown in Fig.(\ref{c}).  Such a critical frequency is equal to
the frequency $m_a/2\pi$ of the radiation produced by axion. In this way, we can determine the axion mass. 
Using the previous study\cite{micro}, we can read the frequency is about $1$GHz to $1.3$GHz although there is
large ambiguity in the data. The previous study did not intended the determination of axion mass.
So in order to derive the mass, the experiment needs to be performed carefully. 

\vspace{0.1cm}
The experiment\cite{sat6} in detail has shown that the saturation arises at temperature below $30$mK. 
When we take into account the correspondence between temperature and frequency of microwave,
we expect that the axion mass is in the range $10^{-5}\mbox{eV}\sim 4\times 10^{-6}$eV.

\begin{figure}[htp]
\centering
\includegraphics[width=0.6\hsize]{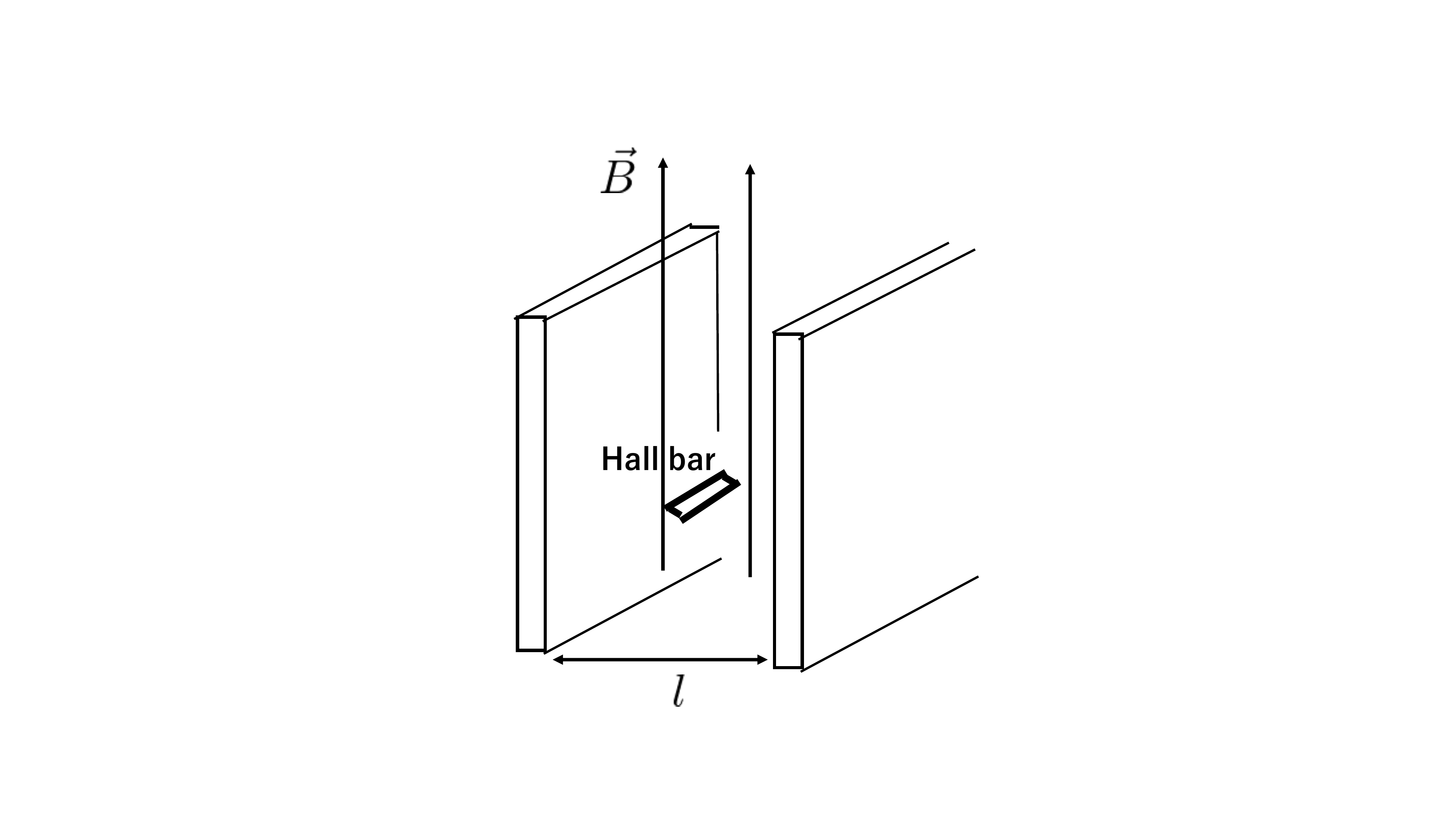}
\caption{Hall bar sandwiched by conducting plates}
\label{d}
\end{figure}

\vspace{0.1cm}
It is favorable to check whether the saturation is really caused by axion. For the purpose, we make radiations by axion much weaker 
and examine whether or not the width diminishes. 
In order to make the radiation weak, we use two parallel conducting plates ( slabs ).
That is, we put the slabs parallel to magnetic field imposed on Hall bar as we have discussed in previous our paper\cite{iwazaki01}. 
Then, the radiation produced by axion changes its strength depending on the position between the slabs.
When we set Hall bar just in the center Fig.\ref{d} between two slabs separated by the distance $l \ll m_a^{-1}$, the strength of the radiation
is suppressed by the factor $(m_al)^2$ 
such that $E_a'\sim g_{a\gamma\gamma}a_0B (m_al)^2/2\sim 10^{-2} (m_a/10^{-5}\mbox{eV})^2(l/0.2\mbox{cm})^2 g_{a\gamma\gamma}a_0B$.   
( This suppression factor can be easily obtained using results in the previous paper\cite{iwazaki01}. )
The spacing $l$ should be much smaller than Compton wave length $1/m_a\sim 2\mbox{cm}(10^{-5}\mbox{eV}/m_a)$.
Consequently, $N_{av}$ is suppressed by the factor $10^{-4}(m_a/10^{-5}\mbox{eV})^4(l/0.2\mbox{cm})^4$ when we choose the spacing $l=0.2$cm.

Additionally, sufficiently large parallel slabs protect radiations generated outside the slabs to penetrate the space between the slabs.
Then, radiations inside the space have only component of electric field parallel to external magnetic field. Such radiations are not 
absorbed in Hall bar because the electric field is perpendicular to two dimensional electrons in the Hall bar. 
Therefore, radiation effect on the saturation of the width 
$\Delta B$ vanishes. So we expect vanishingly small $\Delta B$ even if the Hall bar is not put on the center between the slabs.

In this way, we will see that the width $\Delta B$ may fairly diminish when we put Hall bar in the center between two parallel conducting plates.
The decrease of the width implies that the saturation is caused by axion dark matter.

\section{conclusion}
\label{9}
We have discussed the effect of the axion dark matter on integer quantum Hall effect. 
The axion produces electromagnetic radiation with energy $m_a$ under strong magnetic field used in the experiments.
Electrons in localized states absorb the radiations and transit to effective extended states in the quantum Hall state. 
The electrons in the states carry electric current. Even if a single electron occupies an extended state, the Hall conductivity
or resistance can change its value and forms a new plateau. The probability of the single electron occupying the
extended state at the energy $E_{n\pm}$ is zero in the absence of the axion dark matter 
when Fermi energy $E_f$ is less than the energy $E_{n\pm}$.
The plateau-plateau transition becomes sharp like step function.
On the other hand, the probability is non negligible in the presence of the axion dark matter because the naive estimation shows that 
the number of electrons $N_{av}$ in extended states 
can be of the order of $1$. $N_{av}$ gradually increases from $0$ to $1$ as Fermi energy $E_f$ increases to approach 
the energy $E_{n\pm}$. 
Consequently, the plateau-plateau transition becomes smooth even at $T=0$. 

\vspace{0.1cm}
Furthermore, we discuss how we determine the axion mass with quantum Hall effect.
By observing the saturation of the width $\Delta B$ with changing frequencies of microwaves
imposed on Hall bar,
we can find the axion mass.  Additionally, we propose a way of the confirmation that the axion dark matter really causes the saturation
$\Delta B \neq 0$ even as $T\to 0$. Using two parallel conducting flat plates, we can make radiations generated by axion much weak.
The Hall bar is set in the center between two plates. Then,
the radiation becomes much weak. As a result, we expect that the width $\Delta B$ of the saturation becomes much smaller compared with
the case without two parallel conducting plates.

\vspace{0.2cm}
The author expresses thanks to A. Sawada and Y. Kishimoto for useful comments. He
also expresses thanks to 
members of theory group in KEK for their hospitality.
This work is supported in part by Grant-in-Aid for Scientific Research ( KAKENHI ), No.19K03832.




\begin{thebibliography}{99}
\bibitem{axion1}R. D. Peccei and H. R. Quinn, Phys. Rev. Lett. 38 (1977) 1440.
\bibitem{axion2}S. Weinberg, Phys. Rev. Lett. 40 (1978) 223.
\bibitem{axion3}F. Wilczek, Phys. Rev. Lett. 40 (1978) 279.
\bibitem{Wil}J. Preskill, M. B. Wise and F. Wilczek, Phys. Lett. 120B (1983) 127.
\bibitem{Wil1}L. F. Abbott and P. Sikivie, Phys. Lett. B120 (1983) 133.
\bibitem{Wil2}M. Dine and W. Fischler, Phys. Lett. B120 (1983) 137.
\bibitem{admx}T. Braine et al, Phys. Rev. Lett. 124, (2020) 101303.
\bibitem{carrack}K. Yamamoto, et al. hep-ph/0101200.
\bibitem{haystac}L. Zhong, et al. Phys. Rev. D97 (2018) 092001.
\bibitem{abracadabra}J. L. Ouellet, et al. Phys. Rev. Lett. 122 (2020) 12, 121802.
\bibitem{organ}B. T. McAllister, et al. Phys. Dark Univ. 18 (2017) 67.
\bibitem{madmax}X. Li, et al.  PoS ICHEP2020 (2021) 645.
\bibitem{brass}D. Horns, J. Jaeckel, A. Lindner, A. Lobanov, J. Redondo, and A. Ringwald,  JCAP 04 (2013) 016.
\bibitem{cast}V. Anastassopoulos, et al. Nature Phys. 13 (2017) 584.
\bibitem{sumico}R. Ohta, et al. Nucl. Instr. Meth. A670 (2012) 73.
\bibitem{iwazaki01}A. Iwazaki, PTEP 2022 (2022) 2, 021B01.
\bibitem{fastradioburst1}A. iwazaki, Phys. Rev. D 91 (2015) 2, 023008.
\bibitem{fastradioburst2}A. Iwazaki, Phys. Rev. D 104 (2021) 043022.
\bibitem{magnon}S. Chigusa, T.Moroi and K. Nakayama, Phys. Rev. D101 (2020) 9, 096013.
\bibitem{center}S. Chigusa, et. al., hep-ph/2302.12756.
\bibitem{girvin}R. E. Prange and S. M. Girvin, Springer Verlag, New York(1987).
\bibitem{von}K. v. Klitzing, G. Gorda and M. Pepper. Phys. Rev. Lett. 45 (1980) 494.
\bibitem{aokiando}H. Aoki and T. Ando, Solid State Commun. 38 (1981) 1079.
\bibitem{halperin}B. I. Halperin, Phys. Rev. B 25 (1982) 2185.
\bibitem{joseph1} X. G. Wen and A. Zee, Phy. Rev. B 47, (1993) 2265.
\bibitem{joseph2} Z.F. Ezawa, and A. Iwazaki, Phys. Rev. B 47 (1993) 7295.
\bibitem{joseph3}I. B. Spielman, J. P. Eisenstein, L. N. Pfeiffer, and K. W. West, Phy. Rev. Lett, 84 (2000) 5808. 
\bibitem{sat1}B. Huckestein,  	Rev. Mod. Phys. 67 (1995) 357.
\bibitem{sat2}  D. Shahar, M. Hilke, C. C. Li, D. C. Tsui, S. L. Sondhi, J. E. Cunningham
and M. Razeghi, Solid State Comm. 107 (1998) 19.
\bibitem{sat3}W. Li, C. L. Vicente, J. S. Xia, W. Pan, D. C. Tsui, L. N. Pfeiffer and K. W. West, Phys. Rev. Lett. 102 (2009) 249901.
\bibitem{sat4}X. Wang, et al. Phys. Rev. B 93 (2016) 075307.
\bibitem{sat5}P. Shan, H. Fu, P. Wang, J. Yang, L. N. Pfeiffer, K. W. West and X. Lin, Phys.E 99 (2018) 118.
\bibitem{sat6}C. B. Gudina, Yu. G. Arapov, E. V. Ilchenko, V. N. Neverov, A. P. Savel’ev, S. M. Podgornykh,
N. G. Shelushinina, M. V. Yakunin, I. S. Vasilievskii and A. N. Vinichenko, Semiconductors 52 (2018) 1551.
\bibitem{micro}L. W. Engel, D. Shahar, Q. Kurdak, and D. C. Tsui, Phys. Rev. Lett. 71 (1993) 2638.
\bibitem{localization}P. W. Anderson, Phys. Rev. 109 (1958) 1492.
\bibitem{ono}Y. Ono,  Phys. Soc. Jpn. 51 (1982) 2055.
\bibitem{andouemura}T. Ando and Y. Uemura, J. Phys. Soc. Jpn. 36 (1974) 959.
\bibitem{simulation}T. Ando, J. Phys. Soc. Jpn. 52 (1983) 1740.
\bibitem{topology1}R. B. Laughlin, Phys. Rev. B 23 (1981) 5632.
\bibitem{topology2} D. J. Thouless, M. Kohmoto, P. Nightingale, and M. den Nijs, Phys. Rev. Lett. 49 (1982) 405.
\bibitem{deltaB}Wei, H. P., D. C. Tsui, M. A. Paalanen, and A. M. M. Pruisken, Phys. Rev. Lett. 61 (1988) 1294.
\bibitem{decoherence}P. Mohanty, E. M. Q. Jariwala, and R. A. Webb, Phys. Rev. Lett. 78 (1997) 3366.
\bibitem{aoki1}H. Aoki and T. Ando, Phys. Rev. Lett. 54 (1985) 831.
\bibitem{aoki2}H. Aoki and T. Ando, J. Phys. Soc. Jpn. 54 ( 1985) 2238.
\bibitem{dfsz}M. Dine, W. Fischler and M. Srednicki, Phys. Lett. 104B (1981) 199. 
\bibitem{dfsz1} A. R. Zhitnitsky, Sov. J. Nucl. Phys. 31 (1980) 260.
\bibitem{ksvz}J. E. Kim, Phys. Rev. Lett. 43, (1979) 103. 
\bibitem{ksvz1}M. A. Shifman, A. I. Vainshtein and V. I. Zakharov, Nucl. Phys. B166 (1980) 493.
\bibitem{sikivie}P. Sikivie, Phys. Rev. D32 (1985) 2988.
\bibitem{width}A. Poux, Z. R. Wasilewski, K. J. Friedland, R. Hey, K. H. Ploog, R. Airey, P. Plochocka, and D. K. Maude, Phys. Rev. B 94 (2016) 075411.
\bibitem{andouemura1}T. Ando, Y. Matsumoto and Y. Uemura, J. Phys. Soc. Jpn. 39 (1975) 279. 
\end{thebibliography}
\end{document}